\def\addr#1{{\small\it #1}}

\def\tento#1{\times10^{#1}}
\def\aet#1#2{\approx #1 \tento{#2}}

\def\etal{{\frenchspacing\it et al.}}
\def\ie{{\frenchspacing\it i.e.}}

\def\beq#1{\begin{equation}\label{#1}}
\def\eeq{\end{equation}}
\def\beqa#1{\begin{eqnarray}\label{#1}}
\def\eeqa{\end{eqnarray}}
\def\eq#1{equation~(\ref{#1})}

\def\bfig{\begin{figure}[h] \centerline{\hbox{}}\vfill}
\def\efig{\end{figure}\vfill\newpage}

\def\pp{\noindent\parshape 2 0truecm 13.6truecm 1truecm 12.6truecm}
\def\rf#1;#2;#3;#4 {\par\pp#1, {\it #2}, {\bf #3}, #4. \par}
\def\rg#1;#2;#3;#4;#5 {\par\pp#1, {\it #2}, {\bf #3}, #4 (``#5"). 
\par}
\def\rn{\pp}

\def\K{{\rm K}}
\def\s{{\rm s}}

\def\erg{{\rm erg}}
\def\cm{{\rm cm}}
\def\eV{{\rm eV}}

\def\st{\sigma_t} 
\def\Ob{\Omega_{igm}}  
\def\Oz{\Omega_0}
\def\soz{\sqrt{1+\Oz z}}

\def\etal{{\frenchspacing\it et al.}}
\def\ie{{\frenchspacing\it i.e.}}

\def\izi{\int_0^{\infty}}

\def\expec#1{\langle#1\rangle}

\def\i{x}        % Ionization
\def\crr{\cr\noalign{\vskip 4pt}}

\def\upp{\zeta}
\def\euv{\expec{E_{uv}}}

\def\uvsigfid{\sigma_{18}}

\def\lrec{\lambda_{rec}}  
\def\lci{\lambda_{ci}}
\def\lpi{\lambda_{pi}}
\def\lcomp{\lambda_{comp}}

\def\zion{z_{ion}}

\def\Tpi{T^*}

\def\Te{T_e}  \def\Tp{T_{\gamma}}  \def\DT{\Delta T}
\def\hce{{\cal H}_{ce}}
\def\hcomp{{\cal H}_{comp}}

\def\spose#1{\hbox to 0pt{#1\hss}}
\def\simlt{\mathrel{\spose{\lower 3pt\hbox{$\mathchar"218$}}
     \raise 2.0pt\hbox{$\mathchar"13C$}}}
\def\simgt{\mathrel{\spose{\lower 3pt\hbox{$\mathchar"218$}}
     \raise 2.0pt\hbox{$\mathchar"13E$}}}
\def\simpropto{\mathrel{\spose{\lower 3pt\hbox{$\mathchar"218$}}
     \raise 2.0pt\hbox{$\propto$}}}

\def\i{x}

\documentstyle[11pt,epsf,rotate]{article}

\begin{document}

%%%%%%%%%%%%%%%%%%%%%%%%%%%%%

\begin{titlepage}   % Not numbered.

\noindent

\begin{center}

\vskip0.9truecm
{\bf

DID THE UNIVERSE RECOMBINE?\\
NEW SPECTRAL CONSTRAINTS ON REHEATING\footnote{
Published in {\it ApJ}, {\bf 423}, 529, March 10, 1994.\\
Submitted June 10 1993, accepted September 14.
Available from\\
{\it h t t p://www.sns.ias.edu/$\tilde{~}$max/y.html} 
(faster from the US) and from\\
{\it h t t p://www.mpa-garching.mpg.de/$\tilde{~}$max/y.html} 
(faster from Europe).\\
}
}

\vskip 0.5truecm

Max Tegmark$^1$ 
\&
Joseph Silk$^2$ 

\smallskip
\addr{$^1$Department of Physics, University of California, 
Berkeley, California  94720}\\
\addr{$^2$Departments of Astronomy and Physics, and
Center for Particle Astrophysics, University of California, 
Berkeley, California 94720}\\

\smallskip
\vskip 0.2truecm

\end{center}

\begin{abstract}
One still cannot conclusively assert that the
universe underwent a neutral phase, despite the new COBE FIRAS
limit $y <2.5\times 10^{-5}$ on
Compton $y$-distortions of the cosmic microwave background.
Although scenarios where the very early ($z\sim 1000$) ionization is
thermal (caused by IGM temperatures 
exceeding $10^4$K) are clearly
ruled out, there is a significant loophole for cosmologies with
typical CDM parameters if the dominant ionization mechanism is
photoionization.
If the ionizing radiation has a typical quasar spectrum, 
then the $y$-constraint implies roughly 
$h^{4/3}\Ob \Omega_0^{-0.28}<0.06$ for fully ionized models.
This means that BDM models with $\Omega_0\approx 0.15$ and reionization
at  $z\approx 1000$ are strongly constrained even in this very
conservative case, and can survive the $y$ test only if
most of the baryons form BDM around the reionization epoch. 
\end{abstract}
\end{titlepage}
%%%%%%%%%%%%%%%%%%%%%%%%%%%%%%%%%%%%%%%%%%%%%%

\section{Introduction}

Recombination of the primeval plasma is 
commonly assumed but was by no means inevitable. 
Theories exist that predict early reionization are as diverse as 
those invoking primordial seed fluctuations that underwent early 
collapse and generated sources of ionizing 
radiation, and models involving decaying or annihilating particles. 
The former class includes cosmic strings and 
textures, as well as primordial isocurvature baryon 
fluctuations. The latter category includes baryon 
symmetric cosmologies as well as decaying 
exotic particles or neutrinos.

The Compton $y$-distortion of the cosmic microwave 
background (CBR) provides a unique constraint on 
the epoch of reionization. In view of the extremely 
sensitive recent FIRAS limit of 
$y < 2.5\times 10^{-5}$, we have reinvestigated 
constraints on the early ionization history of the 
intergalactic medium (IGM), and have chosen to 
focus on what we regard as the most important of the 
non-standard recombination history models, namely 
the primordial isocurvature baryon scenario  involving a universe dominated by
baryonic dark matter (BDM), 
as advocated by 
Peebles (1987); Gnedin \&
Ostriker (1992) (hereafter ``GO");  Cen, Ostriker \&
Peebles (1993) and
others.  This class of models takes the simplest matter content 
for the universe, namely baryons, to constitute dark 
matter in an amount that is directly observed and is 
even within the bounds of primordial nucleosynthesis, 
if interpreted liberally, and can reconstruct 
essentially all of the observed phenomena that constrain 
large-scale structure. The BDM model is a non-starter 
unless the IGM underwent very early reionization, in 
order to avoid producing excessive CBR fluctuations 
on degree scales. Fortunately, early nonlinearity is 
inevitable with BDM initial conditions, 
$\delta\rho/\rho\propto M^{-5/12}$, corresponding to 
a power-spectrum $\expec{\delta_k^2}\propto k^{-1/2}$
for the observationally preferred choice of spectral index (Cen,
Ostriker \& Peebles 1993).

Is it possible that the
IGM has been highly ionized since close to the  
standard recombination epoch at $z\approx
1100$? Perhaps the most carefully studied BDM scenario in 
which this happens is that by GO. In their scenario, 
$\Oz=\Omega_{b0}\approx 0.15$. Shortly after recombination, a
large fraction of the mass condenses into faint stars or massive black
holes, releasing energy that reionizes the universe and heats it to
$T>10,000\K$ by $z=800$, so Compton scattering off of hot electrons
causes strong spectral distortions in the cosmic microwave background.
The models in GO give a Compton $y$-parameter between 
$0.96\times 10^{-4}$ and $3.1\times 10^{-4}$, and are thus all ruled
out by the most recent observational constraint from the COBE FIRAS
experiment, $y<2.5\times 10^{-5}$ (Mather {\etal} 1994).

There are essentially four mechanisms that can heat
the IGM sufficiently to produce Compton $y$-distortions:

\begin{itemize}

\item Photoionization heating from UV photons 
(Shapiro \& Giroux 1987; Donahue \& Shull 1991)

\item Compton heating from UV photons 

\item Mechanical heating from supernova-driven winds 
(Schwartz {\etal} 1975; Ikeuchi 1981; Ostriker \& Cowie 1981)

\item Cosmic ray heating 
(Ginzburg \& Ozernoi 1965)

\end{itemize}

\noindent
The second effect tends to drive the IGM temperature towards
two-thirds of the temperature of the ionizing radiation, whereas
the first effect tends to drive the temperature towards a lower value
$T^*$ that will be defined below. The third and fourth
effect can produce
much higher temperatures, often in the millions of degrees. 
The higher the
temperature, the greater the $y$-distortion.

In the GO models, the second effect dominates, which is why they fail so
badly. 
In this paper, we wish to place limits that are
virtually impossible to evade. 
Thus we will use the most cautions assumptions possible, and 
assume that the
latter three heating mechanisms are negligible.

\section{The Compton $y$-Parameter}

Thomson scattering between CBR photons and hot electrons affects the
spectrum of the CBR. It has long been known that hot ionized IGM
causes spectral distortions to the CBR, known as the Sunyaev-Zel'dovich
effect. A useful measure of this distortion is the  Comptonization
$y$-parameter (Kompan\'eets 1957;  
Zel'dovich \& Sunyaev 1969;
Stebbins \& Silk 1986; Bartlett \& Stebbins 1991)
\beq{yDefEq}
y = \int\left({k\Te-k\Tp\over m_e c^2}\right) n_e\st c\> dt
= y^*\int{(1+z)\over\sqrt{1+\Oz z}}\DT_4(z)x(z)dz,
\eeq
where
$$y^* \equiv
\left[1 - \left(1-{x_{He}\over 4x}\right)Y\right] 
\left(k\times10^4\K\over m_e c^2\right)\left({3H_0\Ob\st c\over 8\pi G
m_p}\right) \aet{9.58}{8} h\Ob.$$
Here $\Te$ is the electron temperature, 
$\Tp$ is the CBR temperature,
$\DT_4\equiv(\Te-\Tp)/10^4\K$, $\Omega_{igm}$ is the fraction of 
critical density in 
intergalactic medium,
and $\i(z)$ is the fraction of the hydrogen that
is ionized at redshift $z$. 
Note that we may have
$\Ob\ll\Omega_b$,
{\ie} all baryons may not be in diffuse form.
The integral is to be taken from the reionization epoch to today.
In estimating the electron density $n_e$, 
we have taken the mass fraction of
helium to be  $Y\approx 24\%$ and assumed $x_{He} \approx x$,
{\ie} that helium never becomes doubly ionized and that the fraction
that is singly ionized equals the fraction of hydrogen that is
ionized. The latter is a very crude approximation, but makes a
difference of only $6\%$. 

Let us estimate this integral by making the approximation that the IGM 
is cold and neutral until a redshift $z_{ion}$, at which it suddenly 
becomes ionized, and after which it 
remains completely ionized with a constant temperature $\Te$.
Then for $\zion\gg 1$ and $\Te\gg  z_{ion}\times 2.7\K$ we obtain 
$$y \aet{6.4}{-8} h\Ob\Oz^{-1/2} T_4\> z_{ion}^{3/2},$$
where $T_4\equiv \Te/10^4\K$.
Substituting the most recent observational constraint 
from the COBE FIRAS experiment,
$y < 2.5\times 10^{-5}$ (Mather {\etal} 1994),
into this expression yields
\beq{zionLimitEq}
\zion < 554T_4^{-2/3}\Oz^{1/3}\left({h\Ob\over 0.03}\right)^{-2/3}.
\eeq
Thus the only way to have $\zion$ as high as $1100$ is to have temperatures
considerably below $10^4\K$. In the following
section, we will see to what extent this is possible.

\section{IGM Evolution in the Strong UV Flux Limit}

In this section, we will calculate the thermal evolution of 
IGM for which

\begin{itemize}

\item the IGM remains almost completely ionized at all times,

\item the Compton $y$-distortion is minimized given this constraint.

\end{itemize}

\subsection{The ionization fraction}

In a homogeneous IGM at temperature $T$ exposed to a
density of $\upp$ UV photons of energy {$h\nu>13.6\,\eV$} per proton,
the ionization fraction $\i$ evolves as follows: 
\beq{5IonizationEq}
 {d\i\over d(-z)} = 
{1+z\over\soz}
\left[\lpi(1-\i) + \lci\i(1-\i) - \lrec\i^2\right],
\eeq
where $H_0^{-1}(1+z)^{-3}$ times the rates per baryon for
photoionization, collisional ionization and recombination are given by
\beq{5RateEq}
\cases{
\lpi \aet {1.04}{12} \left[h\Ob\uvsigfid\right]\upp,&\crr
\lci \aet{2.03}{4} h\Ob T_4^{1/2} e^{-15.8/T_4,}&\crr
\lrec \approx 0.717 h\Ob 
T_4^{-1/2}\left[1.808-0.5\ln T_4 + 0.187 T_4^{1/3}\right],&
}
\eeq
and $T_4\equiv \Te/10^4\K$. 
Here $\uvsigfid$ is the spectrally-averaged 
photoionization cross section in units of $10^{-18}\cm^2$.
The differential cross section is given by
(Osterbrock 1974)
\beq{SigmaEq}
{d\uvsigfid\over d\nu}(\nu) \approx 
\cases{
0
&if $\nu < 13.6\,\eV$,\cr
6.30{e^{4-4\arctan(\epsilon)/\epsilon}\over
\nu^4\left(1-e^{-2\pi/\epsilon}\right)}
&if $\nu\ge 13.6\,\eV$,
}
\eeq
where 
$$\epsilon\equiv\sqrt{{h\nu\over 13.6\,\eV}-1}.$$
The recombination rate is the total to all hydrogenic levels
(Seaton 1959; Spitzer 1968). 
Recombinations directly to the ground
state should be included here, since as will become evident below,
the resulting UV photons are outnumbered by the UV photons 
that keep the
IGM photoionized in the first place, and thus can be neglected when
determining the equilibrium temperature.

At high redshifts, the ionization and recombination rates greatly exceed the
expansion rate of the universe, and the ionization level quickly
adjusts to a quasi-static equilibrium value for which the expression in
square brackets in \eq{5IonizationEq} 
vanishes. In the
absence of photoionization, an ionization fraction $x$ close to
unity requires $\Te>15,000\K$. Substituting this into
\eq{zionLimitEq} gives consistency with
$z_{ion}>1000$ only if $h\Ob < 0.008$, a value clearly inconsistent with
the standard nucleosynthesis constraints (Smith {\etal}
1993). Thus any reheating scenario that relies on collisional
ionization to keep the IGM ionized at all times may be considered
ruled out by the COBE FIRAS data.  

However, this does not rule out all ionized universe scenarios,
since photoionization can achieve the same
ionization history while causing a much smaller $y$-distortion.
The lowest temperatures (and hence the smallest
$y$-distortions) compatible with high ionization will be
obtained when the ionizing flux is so strong that 
$\lpi\gg\lci$.
In this limit, to a good
approximation, \eq{5IonizationEq} can be replaced by the
following simple model for the IGM:
 
\begin{itemize}

\item It is completely ionized ($\i=1$).

\item When a neutral hydrogen atom is formed
through recombination, it is instantly photoionized again.

\end{itemize}

\noindent
Thus the only unknown parameter is the IGM temperature $\Te$, which
determines the recombination rate, which in turn equals the
photoionization rate and thus determines the rate of heating.

\subsection{The spectral parameter $\Tpi$}

The net effect of a recombination and subsequent photoionization is to
remove the kinetic energy ${3\over 2}kT$ from the plasma and replace it with
the kinetic energy ${3\over 2}kT^*$, where $\Tpi$ is defined by
${3\over 2}k\Tpi\equiv\euv - 13.6\,\eV$ and $\euv$ is the average energy
of the ionizing photons.
Thus the higher the recombination rate, the faster this effect will tend to
push the temperature towards $\Tpi$. 

The average energy of
the ionizing photons is given by the spectrum $P(\nu)$ as
$\euv = h\expec{\nu}$,
where  $$\expec{\nu} = 
{\izi P(\nu)\sigma(\nu) d\nu\over
 \izi \nu^{-1}P(\nu)\sigma(\nu) d\nu}.$$
Here $\sigma$ is given by \eq{SigmaEq}.
Note that, in contrast to certain nebula calculations where all photons get
absorbed sooner or later, the spectrum should be weighted by the
photoionization cross section. This is because most photons never get
absorbed, and all that is relevant is the energy distribution of those
photons that do. Also note that $P(\nu)$ is the energy distribution
($W/Hz$), not the number distribution which is proportional to 
$P(\nu)/\nu$.

\begin{table}
$$
\begin{tabular}{|llrr|}
\hline
UV source&Spectrum $P(\nu)$&$\euv$&$\Tpi$\\
\hline
O3 star&$T=50,000$K Planck&17.3\,eV&28,300K\\
O6 star&$T=40,000$K Planck&16.6\,eV&23,400K\\
O9 star&$T=30,000$K Planck&15.9\,eV&18,000K\\
Pop. III star&$T=50,000$K Vacca&18.4\,eV&36,900K\\
Black hole, QSO&$\alpha=1$ power law&18.4\,eV&37,400K\\
?&$\alpha=2$ power law&17.2\,eV&27,800K\\
?&$\alpha=0$ power law&20.9\,eV&56,300K\\
?&$T=100,000$K Planck&19.9\,eV&49,000K\\
\hline
\end{tabular}
$$
\caption{Spectral parameters}
\label{ytable1}
\end{table}

The spectral parameters $\euv$ and $\Tpi$ are given in 
Table~\ref{ytable1}
for some selected spectra.
A power law spectrum $P(\nu)\propto
\nu^{-\alpha}$ with $\alpha=1$ fits observed QSO spectra rather well
in the vicinity of the Lyman limit (Cheney \& Rowan-Robinson 1981;
O'Brien {\etal} 1988), and is also consistent with the standard model
for black hole accretion. A Planck spectrum 
$P(\nu)\propto\nu^3/\left(e^{h\nu/kT}-1\right)$
gives a decent prediction of $T^*$ for stars with surface temperatures
below $30,000\K$. For very hot stars, more realistic spectra (Vacca 1993)
fall off much slower above the Lyman limit, thus giving higher values of 
$T^*$. As seen in 
Table~\ref{ytable1},
an extremely metal poor star of surface
temperature $50,000\K$ gives roughly the same $T^*$ as QSO radiation.
The only stars that are likely to be relevant to early
photoionization scenarios are extremely hot and short-lived ones, since 
the universe is less than a million years old at $z = 1000$, and fainter
stars would be unable to inject enough energy in so short a time.
Conceivably, less massive stars could play a the dominant role later on,
thus lowering $T^*$. However, since they radiate such a small fraction of
their energy above the Lyman limit, very large numbers would be needed,
which could be difficult to reconcile with the absence of observations of
Population~III stars today.

\subsection{The thermal evolution}

\begin{figure}[phbt]
\centerline{\epsfxsize=17cm\epsfbox{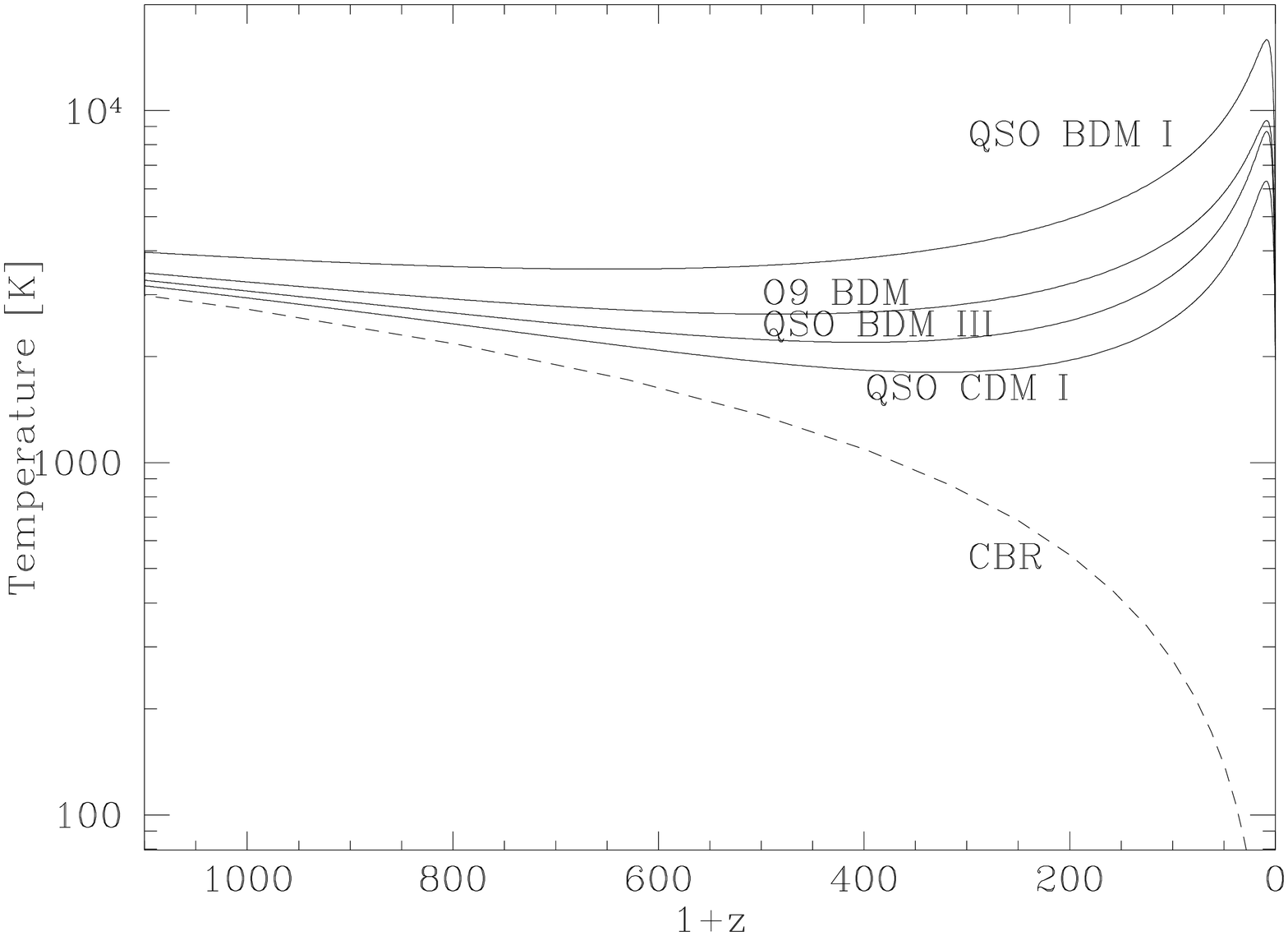}}
\caption{Thermal histories for various models.}
The temperature of the photoionized IGM is plotted for four of the
cosmological models and spectra of ionizing radiation listed in
Table 2. 
The lowermost curve gives the temperature of the CMB photons.
\label{yfig1}
\end{figure}

At the low temperatures involved, the two dominant cooling 
effects\footnote
{
Another cooling mechanism is collisional excitation of atomic hydrogen
followed by radiative de-excitation, which cools the IGM at a rate of
(Dalgarno \& McCray 1972) $$\hce \approx
7.5\tento{-19}  e^{-11.8/T_4} n^2 (1-x)x  \>\erg\, \cm^{-3} \s^{-1}.$$
The ratio of this cooling rate to the Compton cooling rate 
is 
$${\hce\over\hcomp}\approx
{\exp\left[9.2-11.8/T_4\right] (1-x)\over(1+z)T_4} h^2\Ob,$$
a quantity which is much smaller than unity for any reasonable
parameter values when $T < 10^4\K.$  
As will be seen, the temperatures
at $z\approx 1000$ are typically a few thousand K,
which with
$h^2\Ob < 0.1$ and $x>0.9$ renders collisional excitation cooling
more than nine orders of magnitude weaker than Compton cooling. 
Hence
we can safely neglect collisional excitations when computing the IGM
temperature, the reason essentially
being that the temperatures are so low that this process is suppressed
by a huge Boltzman factor.
}
are
Compton drag against the microwave background photons and cooling due to the
adiabatic expansion of the universe. Combining these effects, we obtain the
following equation for the thermal evolution of the IGM:
\beq{5Teq}
{dT\over d(-z)} = -{2\over 1+z}T + 
{1+z\over\soz}\left[\lcomp(\Tp-T) +
{1\over 2}\lrec(T)(\Tpi-T)\right],
\eeq
where
$$\lcomp = {4\pi^2\over 45}
\left({k\Tp\over\hbar c}\right)^4
{\hbar\st\over H_0 m_e}(1+z)^{-3}
\approx 0.00417 h^{-1}(1+z)$$
is $(1+z)^{-3}$ times the Compton cooling rate per Hubble time
and $\Tp = {\Tp}_0(1+z)$.
The factor of ${1\over 2}$ in front of $\lambda_{rec}$ is due to the
fact that the photoelectrons end up sharing their energy with the
protons. We have taken ${\Tp}_0\approx 2.726\K$ (Mather
{\etal} 1994). Numerical solutions to this equation are shown in
Figure~\ref{yfig1},
and the resulting $y$-parameters are given in
Table~\ref{ytable2}.

\begin{table}
$$
\begin{tabular}{|l|rrrrrcl|}
\hline
Model&$\Oz$&$\Ob$&h&$\Tpi$&$z_{ion}$&$y/0.000025$&Verdict\\
\hline
QSO BDM I&0.15&0.15&0.8&37,400K&1100&6.30&Ruled  out\\
QSO BDM II&0.15&0.15&0.8&37,400K&200&1.43&Ruled  out\\
O9 BDM&0.15&0.15&0.8&18,000K&800&2.91&Ruled out\\
QSO BDM III&0.15&0.04&0.8&37,400K&1100&0.67&OK\\
QSO CDM I&1&0.06&0.5&37,400K&1100&0.17&OK\\
QSO CDM II&1&0.03&0.8&37,400K&1100&0.16&OK\\
\hline
\end{tabular}
$$
\caption{Compton $y$-parameters for various scenarios}
\label{ytable2}
\end{table}

\begin{figure}[phbt]
\centerline{\epsfxsize=17cm\epsfbox{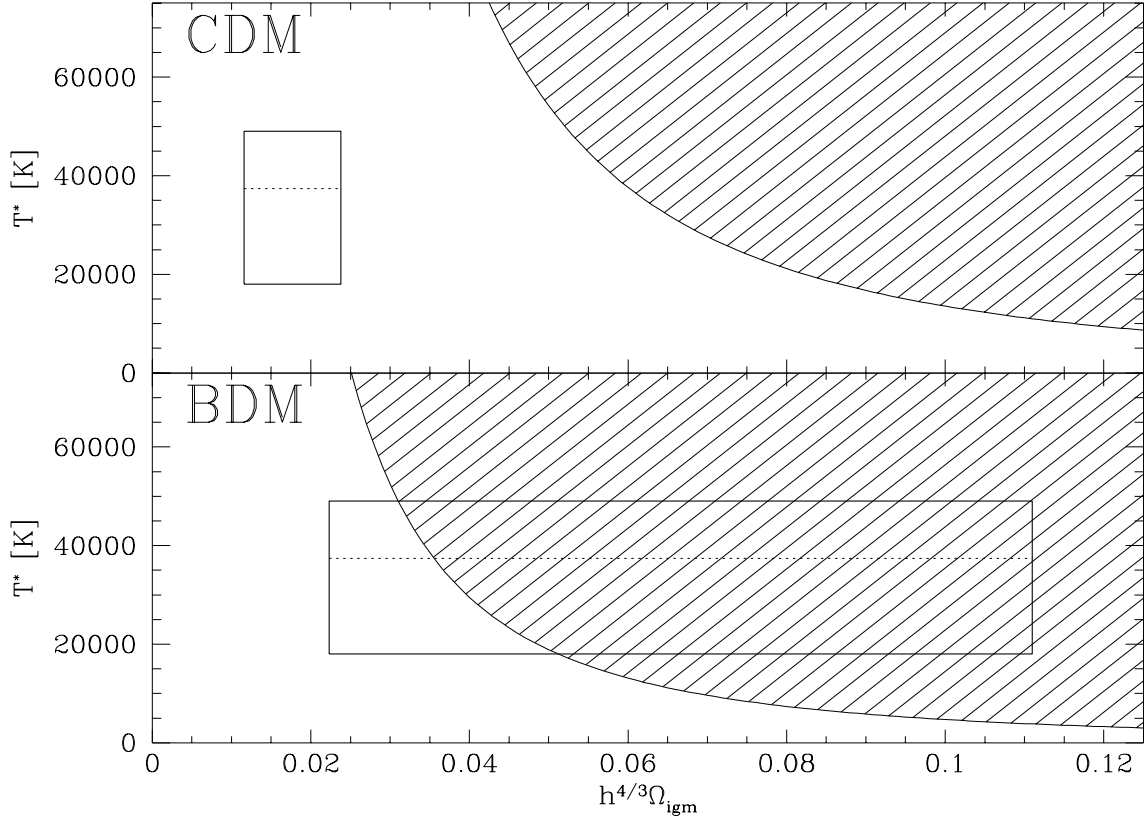}}
\caption{Predicted and ruled out regions of parameter space.}
The hatched regions of parameter space are ruled out by the
the COBE FIRAS limit {$y<2.5\times
10^{-5}$} for $z_{ion}=1100$. $\Omega_0 = 1$ in the CDM plot and
$\Omega_0 = 0.15$ in the BDM plot.
The rectangular regions are the assumed parameter values
for the CDM and BDM models, respectively.
For CDM, the range $0.012 <  h^{4/3}\Omega_{igm} < 0.024$ is given by
the nucleosynthesis constraint $0.010 < h^2\Omega_{b} < 0.015$ and the
assumption that $0.5 < h < 0.8$. (If $\Omega_{igm} < \Omega_b$, the
rectangle shifts to the left.)
For the BDM models, $h=0.8$ and $0.03\leq\Omega_{igm}\leq\Omega_0$.
The vertical range corresponds to
feasible values of the spectral parameter
$T^*$.
The upper limit corresponds to highly speculative star with
surface temperature $100,000\K$ and $T^*=49,000\K$. The lower line
corresponds to an O9 star.
The dotted horizontal line
corresponds to the spectrum expected from quasars/accreting black
holes.
\label{yfig3}
\end{figure}

The temperature
evolution separates into two distinct phases. In the first phase,  
which is almost instantaneous
due to the high recombination rates at low temperatures, 
$T$ rises very rapidly, up to a quasi-equilibrium temperature slightly
above the temperature of the microwave background photons.
After this, in the second phase, $T$ changes only slowly,
and is approximately given by setting the expression
in square brackets in \eq{5Teq}  
equal to zero.
This quasi-equilibrium temperature is typically much lower
than $T^*$, since Compton cooling is so efficient at the high 
redshifts involved, and is given by 
\beq{DTeq}
\DT\equiv\Te-\Tp
\approx {\lrec\over 2\lcomp}(\Tpi-\Te)
\propto  
{1\over 1+z} g(\Te)h^2 \Ob (\Tpi-T),
\eeq
independent of $\Oz$,  
where $g(\Te)\simpropto T^{-0.7}$ encompasses the
temperature dependence of $\lrec$. 
We typically have $T\ll\Tpi$.     
Using this, making the crude approximation 
of neglecting the temperature dependence of $\lrec$,
and substituting \eq{DTeq} into 
\eq{yDefEq}
indicates that    
$$y\simpropto h^3\Ob^2\Omega_0^{-1/2} T_4^* z_{ion}^{1/2}.$$
Numerically selecting the best power-law fit, we find that this is
indeed not too far off: 
the approximation
\beq{yApproxEq}
y\approx 0.0012 h^{2.4}\Ob^{1.8}\Oz^{-1/2}(T^*_4)^{0.8}
(z_{ion}/1100)^{0.9}
\eeq
is accurate to about $10\%$ within the parameter range of cosmological
interest.
We have used \eq{yApproxEq} in
Figure~\ref{yfig3}
by setting  $y=2.5\times 10^{-5}$ and $z_{ion}=1100$. The shaded
region of parameter space is thus ruled out by the COBE FIRAS experiment
for fully ionized scenarios.

\section{Conclusions}

A reanalysis of the Compton $y$-distortion
arising from early reionization
shows that despite the radical sharpening of 
the  FIRAS limit on $y$, 
one still cannot conclusively assert that the universe underwent
a  neutral phase. 
Non-recombining scenarios where the ionization is
thermal, caused by IGM temperatures exceeding $10^4\K$, are clearly
ruled out. Rather, the loophole 
is for the dominant ionization mechanism to be photoionization.
We have shown that for spectra characteristic of both QSO radiation
and massive metal-poor stars, the resulting IGM
temperatures are so low that typical CDM models with no recombination
can still survive the FIRAS test by a factor of six. 
This conclusion is valid if the flux of ionizing radiation
is not so extreme that Compton heating becomes important. This
is not difficult to arrange, as 
the cross section for Thomson scattering is some
six orders of magnitude smaller than that for photoionization.

For BDM models, the constraints are sharper. 
Non-recombining ``classical" BDM models with
$\Omega_{igm}=\Omega_0\approx 0.15$ are ruled out even with the
extremely cautious reheating assumptions used in this paper, the
earliest ionization redshift allowed being  $z \approx 130$.
Such models involving early non-linear 
seeds that on energetic grounds can 
very plausibly provide a photoionization 
source capable of reionizing the universe 
soon after the period of first recombination 
inevitably generate Compton distortions 
of order $10^{-4}$. These include 
texture as well as BDM models, both
of which postulate, and indeed require, 
early reionization ($z > 100$) to 
avoid the generation of excessive 
anisotropy in the cosmic microwave background 
on degree angular scales.

Thus BDM models with reionization at $z\approx 1000$ 
can survive the
$y$ test only if most of the baryons form BDM when reionization
occurs, and are thereby removed as a source of 
$y$-distortion, at least in the diffuse phase.
This may be difficult to arrange
at $z>100,$ since once the matter is reionized  
at this high a redshift, Compton drag is extremely 
effective in inhibiting any further gas collapse until 
$z<100.$ Since it takes only a small fraction 
of the baryons in the universe to provide a source 
of photons sufficient to maintain a fully 
ionized IGM even at $z\sim 1000,$ we suspect 
that most of the baryons remain diffuse 
until Compton drag eventually becomes ineffective. 
Moreover, the possibility that the IGM is only 
partially reionized at $z\sim 1000$ ({\it e.g.} GO), 
a situation which allows a lower value of the  
$y$-parameter, seems to us to be implausible 
as a delicate adjustment of ionization and 
recombination time-scales over a considerable 
range in $z$ would be required.  A 
complementary argument that greatly restricts 
the parameter space allowable for fully 
ionized BDM models appeals to temperature 
fluctuations induced on the secondary last 
scattering surface, both by first order 
Doppler terms on degree scales and by 
second order terms on subarcminute 
scales (Hu {\etal} 1994).
Thus, BDM models would seem to be in 
some difficulty because of the low 
limit on a possible $y$-distortion.

\bigskip
The authors would like to thank
W. Hu, A. Reisenegger and D. Scott for many useful comments, 
and W. Vacca for providing stellar spectra. 
This research has been supported in part by a grant from the NSF.

\clearpage

%%%%%%%%%%%%%%%%%%%%%% REFERENCES: %%%%%%%%%%%%%%%%%%%%%%%%%

\section{REFERENCES}

\rf Arons, J. \& Wingert, D. W. 1972;Ap. J.;177;1

\rf Bartlett, J. \& Stebbins, A. 1991;Ap. J.;371;8

\rn
Cen, R., Gnedin, N. Y., Koffmann, L. A., \& Ostriker, J. P.
1992, Preprint
 
\rn
Cen, R., Ostriker, J. P. \& Peebles, P. J. E 1993, preprint 

\rf Cheney, J. E. \& Rowan-Robinson, M. 1981;MNRAS;195;831

\rf Couchman, H. M. P. 1985;MNRAS;214;137

\rf Couchman, H. M. P. \& Rees, M. 1988;MNRAS;221;53

\rf Dalgarno, A. \& McCray, R. A. 1972;A. Rev. Astr. Astrophys;10;375

\rf Donahue, M. \& Shull, J. M. 1991;Ap. J.;383;511

\rf Feynman, R. P. 1939;Phys. Rev.;56;340

\rf Ginsburg, V. L. \& Ozernoi, L. M. 1965; Astron. Zh.;42;943
(Engl. transl. 196 Sov. Astron. AJ, 9, 726)

\rf Gnedin, N. Y. \& Ostriker, J. P.1992;Ap. J.;400;1

\rn Hu, W., Scott, D. and Silk, J. 1993, preprint.

\rf Ikeuchi, S. 1981;Publ. Astr. Soc. Jpn.;33;211

\rf Kompan\'eets, A. 1957;Soviet Phys. -- JETP;4;730

\rn
Mather {\etal} 1993, preprint.

\rf O'Brien P.T., Wilson, R \& Gondhalekar, P. M 1988;MNRAS;233;801

\rn Osterbrock, D. E. 1974, {\it Astrophysics of Gaseous Nebulae}
(Freeman, San Francisco)
 
\rf Ostriker, J. P. \& Cowie, C. F. 1981; Ap. J.;243;L127

\rf Peebles, P. J. E. 1987; Ap. J.;315;L73

\rf Schwartz, J., Ostriker, J. P., \& Yahil, A. 1975;Ap. J.;202;1

\rf Seaton, M. 1959;MNRAS;119;84

\rf Shapiro, P. R. \& Giroux, M. L. 1987;Ap. J.;321;L107

\rf Smith, M. S., Kawano, L. H. \& Malaney, R. A. 
1993;Ap. J. S.;85;219

\rn Spitzer, L. 1968, {\it Diffuse Matter in Space} (Wiley, New York).

\rn Sugiyama, N. 1993, private communication.

\rf Stebbins, A., \& Silk, J. 1986;Ap. J.;300;1

\rn W. Vacca 1993, private communication.

\rf Zel'dovich, Y., \& Sunyaev, R. 1969;Ap. Space Sci.;4;301

%%%%%%%%%%%%%%%%%%%%%% FIGURES: %%%%%%%%%%%%%%%%%%%%%%%%%

\clearpage

\clearpage
\begin{figure}[phbt]
\centerline{\epsfxsize=17cm\epsfbox{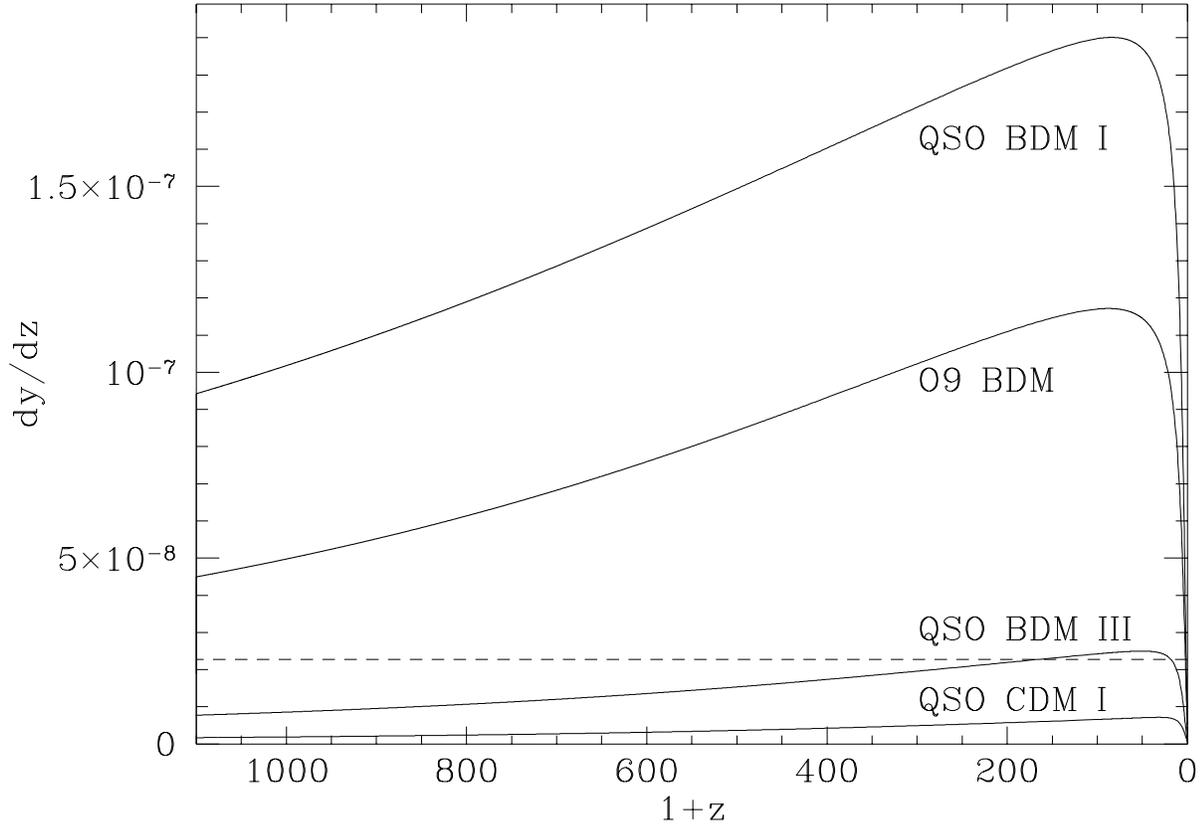}}
\caption{$dy/dz$ for various models.}
The contribution to the y-parameter from different redshifts is
plotted four of the cosmological models and spectra of ionizing
radiation listed in Table 2. Thus for each model, the area under
the curve is the predicted y-parameter. The area under the 
horizontal dashed line is $2.5\times 10^{-5}$, {\it i.e.} the COBE
FIRAS limit.\label{yfig2}
\end{figure}

\clearpage

\end{document}